\documentclass[nofootinbib,showpacs,preprintnumbers,amsmath,amssymb,floatfix]
{revtex4-1}
\usepackage{epsf}

\begin{document}


\title{Factorization and infrared properties of non-perturbative contributions to
 DIS structure functions}

\vspace*{0.3 cm}

\author{B.I.~Ermolaev}
\affiliation{Ioffe Physico-Technical Institute, 194021
 St.Petersburg, Russia}
 \author{M.~Greco}
\affiliation{Department of Physics and INFN, University Roma Tre,
Rome, Italy}
\author{S.I.~Troyan}
\affiliation{St.Petersburg Institute of Nuclear Physics, 188300
Gatchina, Russia}

\begin{abstract}
In this paper we present a new derivation of the QCD factorization.
We deduce the $k_T$- and collinear factorizations
for the DIS structure functions by consecutive reductions of a
more general theoretical construction.
We begin by studying the amplitude of
the forward
Compton scattering off a hadron target, representing this amplitude as a set of convolutions of two blobs
connected by the simplest, two-parton intermediate states.
Each blob in the convolutions
can contain both the perturbative and non-perturbative
contributions.
We formulate
conditions for separating the perturbative and non-perturbative
contributions and attributing them to the different blobs.
After that the convolutions correspond to the QCD factorization.
Then we reduce this totally unintegrated (basic) factorization
first to the $k_T$- factorization and finally to the collinear factorization.
In order to yield a finite expression for the Compton amplitude,
the integration over the loop momentum in the basic
factorization must be free of both ultraviolet
and infrared singularities. This obvious mathematical requirement leads to
theoretical restrictions on the non-perturbative contributions
(parton distributions) to the Compton
amplitude and the DIS structure functions related to the Compton amplitude
through the Optical theorem.
In particular, our analysis excludes
the use of the singular factors $x^{-a}$ (with $a > 0$) in the fits for the quark and gluon distributions
because such factors contradict to
the integrability of the basic convolutions for the Compton amplitude.
This restriction is valid for all DIS structure functions
in the framework of both the $k_T$- factorization and the collinear factorization
if we attribute the
perturbative contributions only to the upper blob.
The restrictions on the non-perturbative contributions obtained in the
present paper
can easily be extended to other QCD processes where the factorization is exploited.
\end{abstract}

\pacs{12.38.Cy}

\maketitle

\section{Introduction}

Factorization is the basic concept for providing the theoretical grounds for the use of perturbative QCD
in the analysis of hadronic reactions.
According to it,
QCD -calculations can be done in two steps: first, the perturbative effects are accounted for with the
calculations of  Feynman graphs involved and second, the perturbative results are complemented
by the appropriate non-perturbative contributions.
There are two basic kinds of the factorization available
in the literature:
the collinear factorization\cite{fact} and $k_T$ -factorization introduced in Ref.~\cite{catani}.
In the framework of the collinear factorization,
the border between those stages is established by introducing
a factorization scale $\mu$
  which at the same time acts as the infrared (IR) cut-off to regulate the
  IR divergences of the perturbative contributions.
This kind of factorization is used in DGLAP\cite{dglap} where
the transverse parton momenta $k_{i~\perp}$ in the perturbative region
are ordered \footnote{The numeration of
$k_{i~\perp}$ in (\ref{dglapord}, \ref{dlord}) runs from the bottom
to the top of the perturbative ladders.} as:
\begin{equation}\label{dglapord}
\mu^2 < k^2_{1~\perp} < k^2_{2~\perp} <... < Q^2~.
\end{equation}
Due to the DGLAP -ordering (\ref{dglapord}), the
virtualities of the initial
partons are smaller  than the virtualities of the ladder partons. The ordering (\ref{dglapord}) and the DGLAP evolution equations were introduced in  the
kinematics where $x = Q^2/2pq \lesssim 1$.\\
In order to obtain a generalization of the DGLAP equations in the small-$x$ region,
the  ordering (\ref{dglapord}) for the virtual quark and gluons in the
perturbative region should be replaced by the ordering obtained in Ref.~\cite{ggfl} :
\begin{equation}\label{dlord}
\mu^2 < \frac{k^2_{1~\perp}}{\beta_1}< \frac{k^2_{2~\perp}}{\beta_2} <... < w = 2pq~.
\end{equation}
We have used in Eqs.~(\ref{dglapord}, \ref{dlord}) the standard notations: $p$ stands for the
moment of the initial hadron, $q$ is the moment of the virtual photon, $Q^2 = -q^2$,
$\beta_r~(r=1,2,..)$ stand for the fractional longitudinal momenta of the ladder partons and the factorization scale
$\mu^2$  is  the starting point of the $Q^2$ -evolution.
The ordering (\ref{dglapord}) is used for accounting for the leading logarithms of $Q^2$
to all orders in $\alpha_s$
while (\ref{dlord}) is used for the total resummation of
all leading logarithms regardless of their arguments. The difference between the orderings
(\ref{dglapord}) and (\ref{dlord}) leads also to different treatments of $\alpha_s$
(see Refs.~\cite{ddt}-\cite{etalfa} for detail).

The common feature of Eqs.~(\ref{dglapord}) and (\ref{dlord}) is that the transverse
momenta $k_{i~\perp}$ are restricted from below. This is necessary in order to
regulate the infrared (IR) divergences arising from the double-logarithmic (DL) contributions.
In this case $\mu$ acts also as an IR cut-off.
Evolving the
scattering amplitudes with respect to $\mu$ is the essence of the Infrared Evolution Equations
(IREE) obtained first in Ref.~\cite{kl} for the quark scattering and then
generalized to various other
high-energy processes (see e.g. Ref.~\cite{acta}).
In particular, a generalization of DGLAP was obtained  in order to
describe
the DIS structure functions $g_1$ and $F_1^{NS}$ at arbitrary $x$ and $Q^2$
(see the overview
of these results in Ref.~\cite{g1sum}).

DL contributions are absent in the BFKL equations\cite{bfkl},
so this approach is IR-stable. As a consequence,
the transverse momenta of the virtual partons here can be arbitrary small.
On the other hand, the initial and final partons (gluons)
in this approach are essentially off-shell. So,
instead of $\mu$, the transverse momenta
 $k_{\perp}$ of the external gluons
 act in BFKL as a new factorization scale, i.e. the border separating the perturbative part from the
 non-perturbative one.
Such a factorization
(the $k_T$ -factorization) was suggested in Ref.~\cite{catani}. The value of
 $k_{\perp}$ is arbitrary, so the $k_T$ -factorization involves the integration over
  $k_{\perp}$. \\
The value of the IR cut-off  $\mu$  is arbitrary (except for the requirement
$\Lambda^2_{QCD} < \mu^2 \ll Q^2$) but
the final expressions for the scattering amplitudes and
structure functions should be insensitive to $\mu$, though
both perturbative and non-perturbative contributions taken alone
depend on it.
The sensitivity of various physical observables to IR cut-offs has been a subject of a great interest
among theorists.
However, such investigations were often focused on the perturbative parts of the observables, leaving
the non-perturbative parts unconsidered
(see e.g. the recent overview \cite{ciaf} and Refs. therein).
In contrast,  we consider in the present
paper the IR- properties of both perturbative and non-perturbative parts.

\begin{figure}[h]
\begin{center}
\begin{picture}(240,180)
\put(0,0){ \epsfbox{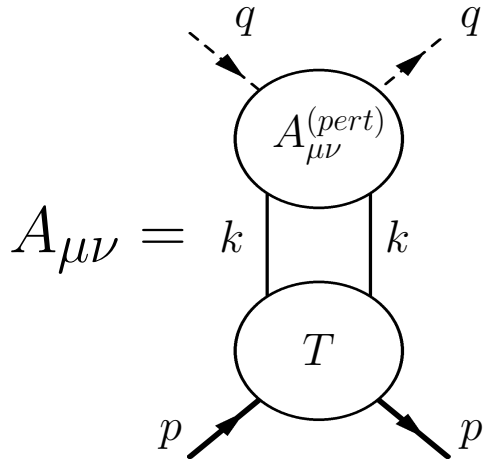} }
\end{picture}
\end{center}
\caption{\label {infraredfig1} Representation of $A_{\mu\nu}$ through the convolution of the perturbative
and non-perturbative blobs.}
\end{figure}

As it is well-known, the factorization means that any scattering
amplitude and any DIS structure function can be represented as a
convolution  of the perturbative and non-perturbative blobs. In
the present paper we consider the photon-hadron (Compton)
scattering and begin with analysis of the forward (at $t = 0$)
Compton amplitude $A_{\mu\nu}$. Representing $A_{\mu\nu}$ as a
convolution of two blobs depicted in Fig.~\ref{infraredfig1}, we
study its integrability and then, using the Optical theorem, we
apply the obtained results to the DIS structure functions.
In contrast, most of the preceding approaches (see e.g. Ref.~\cite{efp})
addressed directly to the DIS hadronic tensor $W_{\mu\nu}$ and
because of that they could not obtain our results and arrive at
the conclusions we make in the present paper. The intermediate
particles in Fig.~\ref{infraredfig1} can be quarks or gluons. The
number of them can be arbitrary. We consider only two-particle
states connecting the
blobs. On one hand, it is the simplest option; on the other hand,
it corresponds (at large $Q^2$) to the leading twist
approximation. All blobs in Fig.~\ref{infraredfig1} are not cut.
The upper blob in Fig.~\ref{infraredfig1} is supposedly obtained
with any perturbative approach, including  DGLAP or BFKL, or IREE
evolution equations, though strictly speaking it can contain also
unperturbative contributions. We study this issue in detail in
Sects.~II,III.

The lower blob $T$ is also a mixture of the perturbative and
non-perturbative contributions. When the radiative corrections to
the upper blob are neglected and the blob is considered in the
Born approximation (see Fig.~\ref{infraredfig2}),  the lower blob
can be regarded as altogether non-perturbative.
For example, this
takes place in the parton model.
  With the QCD radiative
corrections taken into account, both the upper and lower blobs in
Fig.~\ref{infraredfig1} acquire the perturbative contributions.
Nevertheless, following the standard terminology, throughout the
present paper we address the lower (upper) blob as the
non-perturbative (perturbative) blob. When the graph in
Fig.~\ref{infraredfig1} is cut in the $s$ -channel, i.e. when all
$s$- channel intermediate particles are on-shell, the lower blob
$\equiv \Im_s T$ corresponds to the probability to find a
constituent (a quark or a gluon) in the incoming hadron.
Throughout the paper we will address $T$ and $\Im T$ as the
unintegrated parton distributions. When the graph is not cut, we
define $T$ as the result of applying the Dispersion Relations to
$\Im T$. In the present paper we use the Feynman gauge for virtual
gluons. On the other hand, we would like to stress that we
consider the inclusive processes only, which makes us free of the
analysis of the
 rapidity divergencies
arising when the semi-inclusive processes are investigated. This kind of divergences was first
investigated in Refs.~\cite{c,b}. See also recent papers \cite{ch} and references therein.

The available in the literature expressions for
the unintegrated parton distributions were obtained without using
any theoretical grounds. The only argumentation for fixing them
was to fit the experimental data. In the present paper we obtain
certain theoretical restrictions on the distributions following
from the integrability of the convolution in
Fig.~\ref{infraredfig1}, i.e. from the obvious mathematical
requirement that the integration over momentum $k$ must be
convergent. In particular, we show that such restrictions are
compatible with the use of the singular factors $\sim x^{-a}$  in
the fits for the initial parton densities for the singlet
structure function $F_1$, providing $a < 1$, and exclude such
factors in the initial parton densities for the other structure
functions.

Our paper is organized as follows: In Sect.~II we consider the
convolution shown in Fig.~\ref{infraredfig1}. It represents
the forward Compton amplitude $A_{\mu\nu}$ in a factorized form.
We express this amplitude in terms of the invariant amplitudes
and, using the Optical theorem, we relate them in the standard way
to the DIS structure functions. In order to make our reasoning
simpler, we discuss in Sect.~III the integrability of the
convolution in Fig.~\ref{infraredfig1} when the Born approximation
is used for the upper blob. In Sect.~IV we consider the
convolution in Fig.~\ref{infraredfig1} beyond the Born
approximation. In this Sect. we reduce this convolution to the
convolution corresponding to the QCD factorization where the upper
blob is free of non-perturbative contributions. This factorization
involves convolutions with totally unintegrated parton
distributions, so we name it the basic form of factorization and
discuss the similarity and difference between it and the standard
($k_T$ and collinear) factorizations. In Sect.~V we study the
impact of the radiative corrections on the integrability
requirements obtained in Sect.~III. In Sect.~VI we show that the
basic factorization can be reduced to the $k_T$-factorization only
approximately and formulate the condition for such a reduction. In
Sect.~VII we continue studying the $k_T$- factorization and
consider restrictions on the parton distributions. In Sect.~VIII
we reduce the $k_T$- factorization to
 the collinear factorization and
formulate restrictions on the singular factors $\sim x^{-a}$
in the DGLAP fits for the initial parton
 densities for the DIS structure functions following
from the integrability of the basic convolutions.
Finally, Sect.~IX is for concluding remarks.

\section{Representation of the forward Compton amplitude as a convolution}

We start by considering the forward Compton scattering  off a
hadron. The amplitude $A_{\mu\nu}$ for this process includes both
the perturbative and non-perturbative contributions. According to
factorization, $A_{\mu\nu}$ can be represented as the sum of the
convolution shown in Fig.~\ref{infraredfig1} where the
intermediate partons are either quarks or gluons:
\begin{equation}\label{aqg}
A_{\mu\nu} = A^{(q)}_{\mu\nu} + A^{(g)}_{\mu\nu} ,
\end{equation}
with
\begin{eqnarray}\label{convaqg}
A^{(q)}_{\mu\nu}(p,q, S_h) =\int \frac{d^4 k}{(2\pi)^4}
\frac{1}{(k^2 + \imath \epsilon)^2}
\hat{k}\tilde{A}_{\mu\nu}^{(q)}(q,k)\hat{k} T^{(q)}\left(k,p,m_h,S_h\right) \\ \nonumber
A^{(g)}_{\mu\nu}(p,q, S_h) = \int \frac{d^4 k}{(2\pi)^4}
\frac{1}{(k^2 + \imath \epsilon)^2}
\tilde{A}^{(g)}_{\mu\nu\rho\sigma} (q,k) T^{(g)}_{\rho\sigma}\left(k,p,m_h,S_h\right)
\end{eqnarray}
where we have used the superscript $q~(g)$ to mark that the
intermediate partons in Fig.~\ref{infraredfig1} are quarks
(gluons). In Eq.~(\ref{convaqg}) $\tilde{A}^{(q)}_{\mu\nu}$ and
$\tilde{A}^{(g)}_{\mu\nu\rho\sigma}$ stand for the upper blobs in
Fig.~\ref{infraredfig1} and $T^{(q)},~T^{(g)}_{\rho\sigma}$ denote
the lower blobs. Following the conventional notations, we have
used
 the subscripts $\mu$ and $\nu$ for the polarizations of the external virtual photon;
$\rho, \sigma$ mark polarizations of the intermediate gluons; $q, p$ and $k$ are
momenta of the photon, hadron and intermediate virtual partons (quarks or gluons)
respectively, $m_h$ and $S_h$ are the
 hadron mass and spin. In what follows we will focus on
the integrations over $k$, so we skip $m_h$ and $S_h$ in all
subsequent formulae.
Throughout the paper we will keep the standard notations
$Q^2 = - q^2$, $w = 2pq$ and $x = Q^2/w$.

Obviously, even at large $Q^2$ and $w$ the amplitudes
$\tilde{A}_{\mu\nu}^{(q)}, ~\tilde{A}^{(g)}_{\mu\nu\rho\sigma}$
can acquire unperturbative contributions at small $k^2$. In order
to keep these amplitudes completely perturbative, such a soft
region should be excluded from Eq.~(\ref{convaqg}). For example,
in the framework of the collinear factorization it is done by
introducing the factorization scale. In contrast, we will do it
with imposing restrictions on the $k^2$ -dependence of $T^{(q)},
T^{(g)}_{\rho\sigma}$ but even before doing so we will address
$\tilde{A}_{\mu\nu}^{(q)}, ~\tilde{A}^{(g)}_{\mu\nu\rho\sigma}$ as
perturbative objects. Besides them, the convolution in
Eq.~(\ref{convaqg}) involves the amplitudes $T^{(q)}$ and
$T^{(g)}_{\rho\sigma}$ corresponding to the lower blob in
Fig.~\ref{infraredfig1} where non-perturbative contributions are
collected. Strictly speaking, $T^{(q,g)}$ can include perturbative
contributions as well. However, these perturbative terms are quite
similar to the perturbative contents of the amplitudes
$\tilde{A}^{(q)}_{\mu\nu}, \tilde{A}^{(g)}_{\mu\nu\rho\sigma}$.
Throughout the paper we focus on the non-perturbative contents of
$T^{(q)}, T^{(g)}_{\rho\sigma}$, so we address them as a
non-perturbative objects. As it is known, the perturbative
amplitudes $\tilde{A}_{\mu\nu}^{(q)} (q,k),
~\tilde{A}^{(g)}_{\mu\nu\rho\sigma} (q,k)$ can contain
IR-divergent contributions. The virtual parton momentum $k$
regulates these divergences and $k^2$~(or $k^2_{\perp}$) acts as
an IR cut-off. We stress that the convolutions in
Eq.~(\ref{convaqg}), though correspond to a factorization of
$A_{\mu\nu}$ into the blobs, do not coincide with either the
collinear or $k_T$- factorizations. Moreover, they do not imply
any separation of the perturbative contributions from the
non-perturbative ones, which is the fundamental concept of the QCD
factorization. Because of that we will not apply the term
"factorization" to the convolutions in Eq.~(\ref{convaqg}) before
obtaining conditions to separate the perturbative and
non-perturbative contributions. Instead, we call them the primary
convolutions.

The imaginary part (with respect to
$s = (q+p)^2$) of $A_{\mu\nu}$  is related to the hadronic tensor $W_{\mu\nu}$:
\begin{equation}\label{aw}
W_{\mu\nu} = \frac{1}{\pi} \Im_{s} A_{\mu\nu}~.
\end{equation}
In order to simplify the tensor structure of $A_{\mu\nu}$, one can use the
conventional DIS projection operators $P^{(r)}_{\mu\nu}$ enlisted in Appendix B.
Using them, we represent $A_{\mu\nu}$ in terms of invariant amplitudes $A_r$:

\begin{equation}\label{ainv}
A_{\mu\nu} = \sum P^{(r)}_{\mu\nu} A_r .
\end{equation}
Therefore every DIS structure function $f_r$ is related to the invariant Compton amplitude $A_r$:
\begin{equation}\label{awinv}
f_r = \frac{1}{\pi} \Im A_r .
\end{equation}

Obviously, each invariant amplitude $A_r$ can be expressed through the primary convolutions
of the perturbative invariant amplitudes $\tilde{A}^{(q,g)}_r$ and the non-perturbative invariant
amplitudes $T^{q,g}$. Borrowing the terminology of the Regge theory,
we can state that there are the
singlet ($A_S$) and non-singlet ($A_{NS}$) invariant amplitudes contributing to
$A_{\mu\nu}$. They have
the vacuum and non-vacuum quantum numbers in the $t$- channel respectively.
It is important that they exhibit
different behavior with respect to $s$: $A_S$ have an extra power of $s$ compared to $A_{NS}$.
Generically, we can write the primary convolutions (\ref{convaqg}) for them as follows:

\begin{equation}\label{convsns}
A_S = \tilde{A}_S \otimes T_S,~~A_{NS} = \tilde{A}_{NS} \otimes T_{NS} .
\end{equation}
Applying Eq.~(\ref{awinv}) to the singlet amplitude $A_S$ yields the singlet
structure function $F_1$
whereas the non-singlet amplitudes $A_{NS}$ lead to other
structure functions.
We would like to stress that this terminology is not quite
adequate. Indeed,  among these "non-singlets" there are
the amplitudes with imaginary parts
corresponding to the flavor singlet  components of the structure functions
(for example, the singlets $g_{1,2}$).

\section{Compton amplitude $A_{\mu\nu}$ in the Born approximations}

In this Sect. we consider the primary convolutions (\ref{convaqg}) in the simplest
case when the perturbative amplitudes
$\tilde{A}_{\mu\nu}^{(q)},~\tilde{A}^{(g)}_{\mu\nu\rho\sigma}$
in Eq.~(\ref{convaqg})
are calculated in the lowest-order approximation. The convolution
for $A_{\mu\nu}$ is depicted in Figs.~2,3. It is convenient to consider this approximation
for $\tilde{A}_{\mu\nu}^{(q)}$ and $\tilde{A}^{(g)}_{\mu\nu\rho\sigma}$ severally.

\subsection{Lowest-order approximation for $A_{\mu\nu}^{(q)}$ }

\begin{figure}[h]
\begin{center}
\begin{picture}(140,160)
\put(0,0){ \epsfbox{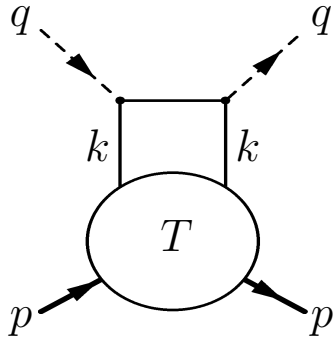} }
\end{picture}
\end{center}
\caption{\label {infraredfig2} Born approximation for the
amplitude of the forward Compton scattering.}
\end{figure}

The case when the amplitude $A_{\mu\nu}^{(q)}$ is calculated in
the Born approximation is depicted in Fig.~\ref{infraredfig2}.

It
corresponds to the use of the parton model. Dropping the QED
quark-photon coupling $e^2_q$ as an unessential factor,
 we can write the Born perturbative contribution
$\tilde{A}_{\mu\nu}^{(q~B)}$ as follows:
\begin{equation}\label{aqborn}
\tilde{A}_{\mu\nu}^{(q~B)} = \gamma_{\nu}\frac{1}{\hat{k} + \hat{q}
+ \imath \epsilon}\gamma_{\mu}
+  \gamma_{\mu}\frac{1}{\hat{k} - \hat{q} + \imath \epsilon}\gamma_{\nu}
\end{equation}

where we have neglected the quark mass.
Therefore, $\hat{k} \tilde{A}_{\mu\nu}^{(q~B)} \hat{k}$ can be written as follows:
\begin{equation}\label{kaqbornk}
\hat{k} \tilde{A}_{\mu\nu}^{(q~B)} \hat{k} = \frac{B_{\mu\nu}(q,k)}{(k+q)^2 + \imath \epsilon}
+ \frac{B_{\nu \mu}(q,k)}{(k-q)^2 + \imath \epsilon} .
\end{equation}

The factors
$B_{\mu\nu}(q,k)$ in Eq.~(\ref{kaqbornk})
are considered in detail in Appendix~A. Before performing
a detailed calculation, let us give a simple
Euclidean-like estimate for the ultraviolet behavior of
$T^{(q)}$ in Eq.~(\ref{convaqg}) when $\tilde{A}^{(q)}_{\mu\nu} = $ is given by its Born
value.
Obviously, we can write
$d^4 k = k^3 d k d \Omega_4$. Then, at large $k$ the Born amplitude in Eq.~(\ref{aqborn})
$\tilde{A}_{\mu\nu}^{(q~B)} \sim 1/k^3$ and therefore at large $k$

\begin{equation}\label{aborneucl}
A^{(q)}_{\mu\nu} \sim \int d k \frac{k^3}{k^3} T^{(q)}(k) .
\end{equation}
This integral is convergent at large $k$ only if
\begin{equation}\label{tqbeuk}
T^{(q)} \sim k^{-1 - h} ,
\end{equation}
with $h > 0$ .

Now let us investigate this case more carefully.
It is convenient to use the Sudakov variables introduced in Ref.~\cite{sud}.
We will use them in the following form
\footnote{We account for $Q^2$ but neglect the quark mass. }:
\begin{equation}\label{sud}
k = -\alpha q' + \beta p + k_{\perp} ,
\end{equation}
with $q' = q + x p$, $x = Q^2/w$, so that $q'^2 \approx 0$ and
\begin{equation}\label{kq}
k^2 = -w \alpha \beta - k_{\perp}^2,~2pk = - w \alpha,~2qk = w(\beta + x \alpha),~
(q+k)^2 = w (\beta - x)(1 - \alpha) - k^2_{\perp}.
\end{equation}

$B_{\mu\nu}$
can be simplified
with using the projection operators $P^{(r)}_{\mu\nu}$
defined in Appendix B.  Then we arrive at the following expressions for
$A_{\mu\nu}^B$:

\begin{equation}\label{convaqborn}
A_{\mu\nu}^{(q)} = \sum_r P^{(r)}_{\mu\nu} A_r^{B} = \sum_r P^{(r)}_{\mu\nu} \int d k^2_{\perp} d \beta d \alpha
\tilde{A}^{(q~B)}_r (q,k)\frac{B}
{(w \alpha\beta + k^2_{\perp} - \imath \epsilon)^2} T^{(q)}_r(k,p)~,
\end{equation}
where $\tilde{A}^{(q~B)}_r$ are the invariant amplitudes in the Born
approximation and $B$ is defined in Eq.~(\ref{kkint}).
We have also provided the non-perturbative amplitudes $T^{q}$ with the superscript $r$ as
they can be different for different $A_rT^{q}$.
Obviously, $T^{(q)}_r(k,p)$ depend on the invariant
energy and the external virtualities:
\begin{equation}\label{tqborn}
T^{(q)}_r(k,p) = T^{(q)}_r((k+p)^2, k^2)=
T^{(q)}_r \left(w\alpha, (w \alpha\beta + k^2_{\perp})\right) .
\end{equation}

The Born invariant amplitudes $\tilde{A}^{(q~B)}_r$ are well-known. For example, the amplitudes
$\tilde{A}^{(q~B)}_1$ and $\tilde{A}^{(q~B)}_3$ related to the structure functions $F_1$ and $g_1$ respectively are
\begin{eqnarray}\label{aqinvborn}
\tilde{A}^{(q~B)}_1  = \left[ \frac{w(1 - \alpha)}
{w \beta - Q^2 - w \alpha\beta - k^2_{\perp}+ \imath \epsilon} +
\frac{-w(1 - \alpha)}
{- w \beta -  Q^2 - w \alpha\beta - k^2_{\perp}+ \imath \epsilon}\right] , \\ \nonumber
\tilde{A}^{(q~B)}_3 = \left[ \frac{w(1 - \alpha)}
{w \beta - Q^2 - w \alpha\beta - k^2_{\perp}+ \imath \epsilon} -
\frac{-w(1 - \alpha)}
{- w \beta - Q^2 - w \alpha\beta - k^2_{\perp}+ \imath \epsilon }\right] .
\end{eqnarray}
Taking the imaginary part of $\tilde{A}^{(q~B)}_1,~\tilde{A}^{(q~B)}_3$ with respect to $w$
and neglecting the virtuality $k^2 = - w \alpha\beta - k^2_{\perp}$, we arrive at
the well-known result of the parton model:
$\Im \tilde{A}^{(pert~B)}_{1,3} \sim  \delta (\beta - x)$.

Now let us integrate  Eq.~(\ref{convaqborn}) with
respect to $\alpha$ and focus on
the region of large $|\alpha|$.
In this region the amplitudes $\tilde{A}_r^{q~B} \sim \alpha/\alpha$,
both the denominator and $B$ in Eq.~(\ref{convaqborn}) are $\sim \alpha^2$,
so the $\alpha$-integration looks at large $|\alpha|$ as follows:

\begin{equation}\label{alphabigbornsud}
\int d \alpha \frac{\alpha^3}{\alpha^3} T^{(q)}_r(\alpha)
\end{equation}
This integral is convergent at large $|\alpha|$ only if
\begin{equation}\label{tqb}
T^{(q)}_r \sim \alpha^{-1 - h} ,
\end{equation}
 with  $h >0$. Eq.~(\ref{tqb}) confirms the estimate
made in Eq.~(\ref{tqbeuk}). According to Eq.~(\ref{sud}), the invariant energy $s' = (p-k)^2$ of $T^{(r)}$
 at large $\alpha$ is $s' = w \alpha - w\alpha\beta - k^2_{\perp} \approx w \alpha$.
 So the meaning of Eq.~(\ref{tqb}) is that $T^{(q)}_r (s')$
should decrease faster than $1/s'$ with growth of $s'$ in order to prevent UV divergency in
Eq.~(\ref{convaqborn}). Obviously, $\tilde{A}^{(q~B)}_1$ has non-vacuum quantum numbers in
the $t$- channel and therefore it contributes to the non-singlet part of the
Compton amplitude $A_{\mu\nu}$.

\subsection{Lowest-order approximation for $A_{\mu\nu}^{(g)}$ }

\begin{figure}[h]
\begin{center}
\begin{picture}(380,160)
\put(0,0){ \epsfbox{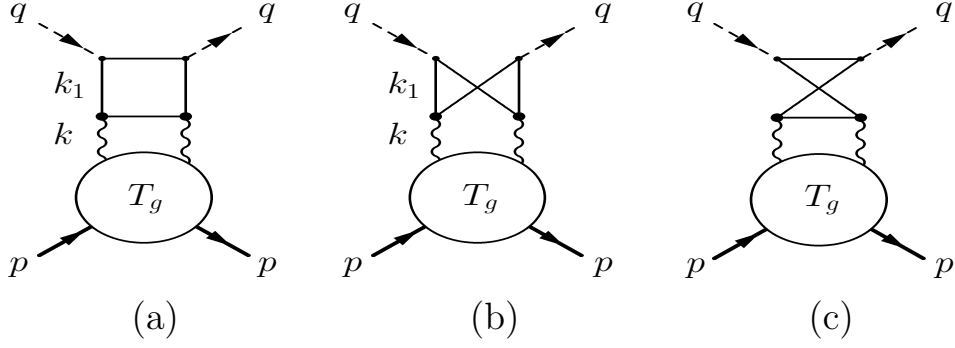} }
\end{picture}
\end{center}
\caption{\label {infraredfig3} Gluon contribution to the forward
Compton scattering in the lowest order.}
\end{figure}

The case when the amplitude $A_{\mu\nu}^{(g)}$ is calculated in
the lowest order is depicted in Fig.~\ref{infraredfig3}.

For the shortness reason, we will address it as the Born
approximation as well.  The primary convolution for the Compton
amplitude $A^{(g)}_{\mu\nu}$ in this case is

\begin{equation}\label{agborn}
A_{\mu\nu}^{(g)} (q,k) = \int \frac{d^4 k}{(2 \pi)^4}
\Gamma_{\mu\nu~\rho\sigma} (q,k)
\frac{1}{\left[k^2 + \imath \epsilon\right]^2} T^{(g)}_{\rho\sigma} (k,p)
\end{equation}
where we have used the notation $\Gamma_{\mu\nu~\rho\sigma} (q,k)$
for the upper, perturbative blob given by the quark boxes, one of
them depicted in Fig.~\ref{infraredfig3}. After applying the
projection operators, Eq.~(\ref{agborn}) can generically be
rewritten in terms of the invariant amplitudes as follows:

\begin{equation}\label{invagborn}
A^{(g)} (q,p) = \int \frac{d^4 k}{(2 \pi)^4} \Gamma_{\rho\sigma} (q,k)
\frac{1}{\left[k^2 + \imath \epsilon\right]^2} T^{(g)}_{\rho\sigma} (k,p) .
\end{equation}

The rough estimate for $T^{(g)}_{\rho\sigma}$ can be obtained similarly to
the one in Eq.~(\ref{tqbeuk}). Writing $d^4 k = k^3 dk d \Omega_4$
and assuming that $k$ is large leads to

\begin{equation}\label{agbeuc}
A^{(g)} \sim \int \frac{d k}{k} T^{(g)}_{\rho\sigma} (k,p) .
\end{equation}
The integration over $k$ is free of the UV singularity when
\begin{equation}\label{tgeucl}
 T^{(g)}_{\rho\sigma} \sim k^{-h} ,
\end{equation}
which differs from Eq.~(\ref{tqbeuk}).

Now let us make a more detailed estimate of the same Born case,
using the Sudakov parametrization (\ref{sud}) for $k$. At large
$\alpha$ Eq.~(\ref{invagborn}) is

\begin{equation}\label{agb}
A^{(g)} \sim \int d \alpha \Gamma (q,k)_{\rho\sigma}
\frac{1}{\alpha^2} T^{(g)}_{\rho\sigma} (\alpha, k^2) .
\end{equation}

The contribution $\alpha^2$ in Eq.~(\ref{agb}) comes from $k^2$.
According to Eq.~(\ref{sud}), $k^2 = -w\alpha \beta - k^2_{1
\perp}$, i.e. $k^2_1 \sim \alpha$ at large $\alpha$.  The
perturbative factor $\Gamma_{\rho\sigma}$  is known to be free of
the UV singularities and can be calculated by integrating over
momentum $k_1$ (see Fig.~\ref{infraredfig3}). This integration
yields different results for the singlet and non-singlet. Also a
dependence of $T^{(g)}_{\rho\sigma}$ on $\alpha$ can be different
for the singlet and non-singlet.

In the singlet case the polarizations $\rho$ and $\sigma$
are longitudinal and $T^{(g)}_{\rho\sigma}$
can be represented as follows:
\begin{equation}\label{tgs}
T^{(g)}_{\rho\sigma} = \frac{2p_{\rho}p_{\sigma}}{w} T^{(g)}_S(\alpha, k^2) ,
\end{equation}
so the integration over $\alpha$
in Eq.~(\ref{invagborn}) is reduced to

\begin{equation}\label{invagalpha}
A^{(g)}_S \sim \int d \alpha \frac{\alpha}{\alpha^2} T^{(g)}_S (\alpha, k^2) .
\end{equation}

This integral is convergent when
\begin{equation}\label{tgbs}
T^{(g)}_S \sim \alpha^{-h} .
\end{equation}

It coincides
with the rough integrability requirement in Eq.~(\ref{tgeucl}) but differs from Eq.~(\ref{tqbeuk})
for the quark contribution.

The non-singlet amplitudes $A_{NS}$ are either flavor non-singlets or spin-dependent. In the
latter case $T^{(g)}_{\rho\sigma}$
is antisymmetrical in $\rho\sigma$ and can be represented as
\begin{equation}\label{gluontns}
T^{(g)}_{\rho\sigma} = \imath \frac{m_h}{w} \epsilon_{\rho\sigma \lambda\tau}
S_{\lambda} k_{\tau}  T^{(g)}_{NS} (\alpha, k^2),
\end{equation}
with $S_{\lambda}, m_h$ being the spin of the hadron target and the mass scale
respectively,
or in a similar form.
Therefore, in the non-singlet case Eq.~(\ref{tgbs}) is replaced by

\begin{equation}\label{gluonalphans}
\int d \alpha \frac{\alpha^2}{\alpha^2} T^{(g)}_{NS} (\alpha, k^2) .
\end{equation}
This integral is convergent when
\begin{equation}\label{tgb}
T^{(g)}_{NS} \sim \alpha^{-1-h} ,
\end{equation}
which coincides with the integrability requirement for the quark non-singlet
case. Therefore, the integrability requirements (\ref{tqb},\ref{tgb}) for the non-singlet
part of $A_{\mu\nu}$ in the lowest-order
approximation do not depend on the kind of the intermediate partons.

\section{Compton amplitudes $A_S,~A_{NS}$ beyond the Born approximation}

 When the perturbative amplitudes $\tilde{A}_S,~\tilde{A}_{NS}$ in Eq.~(\ref{convsns})
are calculated in the Born approximation, the amplitudes $T_S,~T_{NS}$ can be
regarded
as completely non-perturbative objects.
Studying $A_S,~A_{NS}$ beyond the Born approximation
brings the radiative corrections  to $\tilde{A}_S,~\tilde{A}_{NS}$.
In principle, the radiative corrections can also be placed into  $T_S,~T_{NS}$.
It converts the Born invariant amplitudes $\tilde{A}^{(q~B)}_{1,3}$ of Eq.~(\ref{aqinvborn})
and $\Gamma_{\rho\sigma}$ of Eq.~(\ref{agborn}) into more involved
invariant amplitudes which we generically denote $A^{(q,g)}_{S,NS}$.
This can have an impact on the $\alpha$-dependence
of the non-perturbative blobs $T^{q,g}_{S,NS}$ and violate the integrability
requirements for the singlet (\ref{tgbs}) and non-singlet (\ref{tqb},\ref{tgb}) amplitudes.
In the previous Sect. we showed that
the integrability requirements do not depend on the kind of the
intermediate partons in the convolutions, so in the present Sect. we
focus on the quark amplitudes $A^{(q)}_S$ and
will skip the superscripts $q,g$ in the sequential formulae.
Each of $A_{S,NS}$ can depend on $\alpha$ through $k^2$ only. Besides, they
include the
infrared-sensitive
radiative corrections becoming singular at
small $k^2$.
In particular, there are
the logarithmic terms $\sim \ln^n (w\beta/k^2), \ln (Q^2/k^2)$.
In addition to them,
the first-loop radiative corrections to the singlet unpolarized
amplitudes $A_S$
yield the infrared-sensitive power term
$w \beta/k^2$.
As it is known, such a term is originated by the
2-gluon intermediate state, with the gluons having the
longitudinal polarizations.
Such amplitudes can be written as follows:

\begin{equation}\label{ams}
\tilde{A}_S = (w\beta/k^2) M_S (\ln (w\beta/k^2), \ln (Q^2/k^2)) .
\end{equation}

As written explicitly, $M_S$  in Eq.~(\ref{ams}) includes the infrared-sensitive
logarithms. Of course, it can also include other, infrared-insensitive corrections.
The non-singlet amplitudes  $A^{q,g}_{NS}$ do not have the power term $(w\beta/k^2)$, so
we write them in the following way:
\begin{equation}\label{apert}
\tilde{A}_{NS}  = M_{NS} (\ln (w\beta/k^2), \ln (Q^2/k^2)) .
\end{equation}

In these terms we can rewrite the primary convolutions in Eq.~(\ref{convsns})
as follows:

\begin{equation}\label{convs}
A_S = \int d k^2_{\perp}\frac{d \beta}{\beta} d \alpha
\left(\frac{w \beta}{k^2} \right)
M_S \left(\ln (w\beta/k^2), \ln (Q^2/k^2)\right) \frac{B}{(w\alpha\beta + k^2_{\perp})^2}
T_S (w \alpha, k^2)
\end{equation}
and

\begin{equation}\label{convns}
A_{NS} = \int d k^2_{\perp}\frac{d \beta}{\beta} d \alpha
M_{NS} \left(\ln (w\beta/k^2), \ln (Q^2/k^2) \right)\frac{B}{(w\alpha\beta + k^2_{\perp})^2}
T_{NS} (w \alpha, k^2) .
\end{equation}

We remind that each of the amplitudes $M_{S,NS}$ and $T_{S,NS}$
in the convolution (\ref{convs},\ref{convns}) can contain both perturbative and
non-perturbative contributions, so these convolutions (we have addressed them as the
primary convolutions) cannot be identified with the factorization convolutions.

\subsection{Converting primary convolutions into factorization convolutions}

$M_S,~M_{NS}$ in Eqs.~(\ref{convs},\ref{convns}) can be entirely perturbative objects only if
the region of small $k^2$ is excluded from these convolutions. Besides, the integrals
in  (\ref{convs},\ref{convns}) should be IR- stable.
These goals can be achieved when

\begin{equation}\label{tnsk2}
T_{NS} \sim (k^2)^{\eta}
\end{equation}
and

\begin{equation}\label{tsk2}
T_S \sim (k^2)^{1 + \eta}~,
 \end{equation}
with $\eta > 0$, at small $k^2$ .

\begin{figure}[h]
\begin{center}
\begin{picture}(280,140)
\put(0,0){ \epsfbox{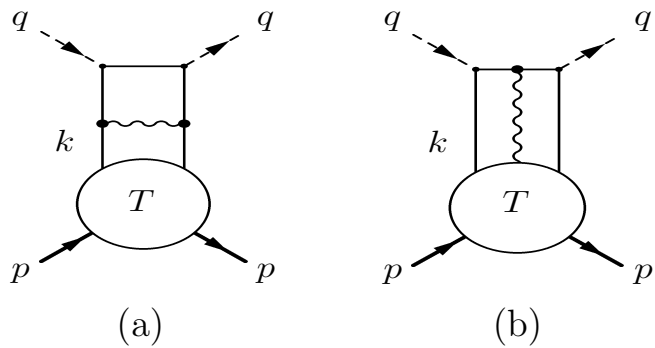} }
\end{picture}
\end{center}
\caption{\label {infraredfig4} Radiative corrections to the Born
amplitude.}
\end{figure}

Imposing conditions (\ref{tnsk2},\ref{tnsk2}) makes the amplitudes
$M_{S,NS}$ (i.e. the upper blob in Fig.~\ref{infraredfig1}) free
of non-perturbative contributions and it also solves the problem
of the mass singularities\cite{masssing}. However the lower blobs,
$T_{S,NS}$ can include both the perturbative and non-perturbative
contributions. When the perturbative blobs were considered in the
Born approximation, the lower blobs were regarded as an absolutely
non-perturbative objects. In this paper we advocate the point of
view that beyond the Born approximation the lower blobs
contain non-perturbative contributions only. Indeed, let us add a
radiative correction (for instance, a ladder gluon propagator as
shown in Fig.~\ref{infraredfig4}a) to the Born graph.

Obviously, it can be included into the upper blob. This procedure
can be repeated each time when a new radiative correction is added
to Figs.~2,3, except for the case when the new propagators connect
 the upper and lower blobs as shown in Fig.~\ref{infraredfig4}b.
Such graphs involve the $t$-channel intermediate states with three or more partons.
Analysis of them requires a special attention (see e.g.
Ref.~\cite{efr}). We do not consider them in the present paper.

Eqs.~(\ref{convs},\ref{convns}) complemented by
the IR-restrictions (\ref{tnsk2}, \ref{tsk2}) rigorously
correspond to the QCD factorization concept
where the non-perturbative
contributions should be attributed to $T_s, T_{NS}$ while the amplitudes $M_S,~M_{NS}$
can be regarded as entirely perturbative objects.
Nevertheless, it is clear that the form of the
convolutions in these equations does not
correspond to the form of either the $k_T$- or the collinear factorization.
Indeed, these factorizations  involve
the same objects (the perturbative and non-perturbative blobs)
as in Eqs.~(\ref{convs},\ref{convns}),
however with the smaller number of the integrations.
Let us remind that
the expressions with the $k_T$-factorization
include two integrations only:
over $\beta$ and $k_{\perp}$, while the collinear factorization
involves the integration over the longitudinal variable
$\beta$ only. In contrast, Eqs.~(\ref{convs},\ref{convns})
additionally include the integration over $\alpha$. In order to
distinguish
the convolutions in Eqs.~(\ref{convs},\ref{convns}) from the convolutions
with the $k_T$- and collinear factorizations, we will call the
totally unintegrated
form of the factorization convolutions in Eqs.~(\ref{convs},\ref{convns}) the basic form
of factorization.
Below we show how to proceed from the basic
form of factorization to
the $k_T$- and collinear factorizations but before it, let us
write down the factorization expressions for the DIS structure functions.

\subsection{Basic form of factorization for the structure functions}

Applying Eq.~(\ref{awinv}) to the invariant Compton amplitudes  (\ref{convns})
complemented by
the IR-restrictions (\ref{tnsk2}, \ref{tsk2}), we arrive at the following expressions for the non-singlet
structure functions in the basic form of factorization:

\begin{equation}\label{convnsf}
f_{NS} = \int d k^2_{\perp}\frac{d \beta}{\beta} d \alpha
f_{NS}^{(pert)} \left(\ln (w\beta/k^2), \ln (Q^2/k^2) \right)\frac{B}{(w\alpha\beta + k^2_{\perp})^2}
\Omega_{NS} (w \alpha, k^2)
\end{equation}
where  $f^{(part)}_{NS} = (1/\pi) \Im M_{NS}$ stands for the perturbative  contributions
 to the non-singlets and $\Omega_{NS} (w \alpha, k^2) \equiv
 \Im T_{NS} (w \alpha, k^2)$ are the totally unintegrated non-singlet
 parton distributions. Similarly, we obtain the following representation for the singlet structure
 function from (\ref{convs}):
\begin{equation}\label{convsf}
f_{S} = \int d k^2_{\perp}\frac{d \beta}{\beta} d \alpha
f_{S}^{(pert)} \left(\ln (w\beta/k^2), \ln (Q^2/k^2) \right)\frac{B}{(w\alpha\beta + k^2_{\perp})^2}
\Omega_s (w \alpha, k^2) ,
\end{equation}
where the perturbative singlet contribution is
\begin{equation}\label{fspert}
f_{S}^{(pert)} = \left(\frac{w\beta}{k^2}\right) (1/\pi) \Im M_{S}
.
\end{equation}
and $\Omega_s (w \alpha, k^2) \equiv \Im T_{S} (w \alpha, k^2)$
stands for the totally unintegrated singlet parton distributions.

\section{Conditions of integrability for non-singlet Compton amplitudes}

Let us consider the integrations in Eq.~(\ref{convns}) for the non-singlet Compton amplitude.
Generally, the order of the integrations in Eqs.~(\ref{convs},\ref{convns}) can be arbitrary. It is convenient to
integrate over $\alpha$  first.
This is usually performed by using
the Cauchy theorem.
So, the integration line $-\infty < \alpha < \infty$ is
complemented by a semi-circle in the upper or lower half of the complex
$\alpha$ -plane and after that singularities of the integrand with in the $\alpha$ -plane
(poles and cuts) should be found. The integral over the semi-circle $C_R$
vanishes when the radius of the semi-circle
tends to infinity, so the result is determined by either residues at the poles or integrals along
the cuts  of the integrand in the $\alpha$-plane.
Now let us consider the integral over $\alpha$ in Eq.~(\ref{convns}).
For integration along $C_R$, where  $|\alpha| \to \infty$, the $\alpha^2$
-terms both in $B(k)$ and in the denominator of Eq.~(\ref{psi})
cannot be neglected, so the integral along $C_R$ behaves as
\begin{equation}\label{alphabigborn}
\int d \alpha \frac{\alpha^3}{\alpha^3} T_{NS}(\alpha, k^2_{\perp}) .
\end{equation}

It vanishes at $|\alpha| \to \infty$ if
\begin{equation}\label{tnsalpha}
T_{NS} \sim \alpha^{-1 - h}
\end{equation}
at large $\alpha$, with arbitrary  $h >0$ .
This restriction coincides with  Eq.~(\ref{tgb})
obtained in the Born approximation. Therefore,
accounting for the radiative corrections to the non-singlet amplitudes does not
change the $\alpha$ -dependence of $T$ at large $|\alpha|$ and has the same meaning:
$T_{NS}$ should decrease faster than $1/s'$ when the invariant energy $s' = (p-k)^2$ grows.

\subsection{Integrability of the singlet Compton amplitude}

The singlet invariant amplitude $A_S$ is defined in Eq.~(\ref{convs}).
The main difference between  Eqs.~(\ref{convs}) and the non-singlet amplitudes of Eqs.~(\ref{convns})
is the factor $w\beta/k^2$ in Eq.~(\ref{convs}).
Therefore, the integration over $\alpha$ in Eq.~(\ref{convs})
leads to new requirements for the singlet amplitude $T_S$.  Using the same arguments as for
$T_{NS}$, we  conclude that the integration over $\alpha$ at $|\alpha| \to \infty$ looks now as follows:

\begin{equation}\label{alphabigs}
\int d \alpha \frac{\alpha^2}{\alpha^3} T_S(\alpha, k^2) .
\end{equation}
This integral is convergent when
 \begin{equation}\label{tsalpha}
T_S \sim \alpha^{- h} ,
\end{equation}
which coincides with the integrability requirement (\ref{tgbs}) for the singlet amplitude
in the lowest-order approximation.
It follows from Eq.~(\ref{sud}), that $w\alpha = 2pk$ is at large $\alpha$
the invariant energy of the
non-perturbative blobs $T_{S,~NS}$, so Eq.~(\ref{tsalpha}),
similarly to Eq.~(\ref{tnsalpha}),
predicts the dependence of $T_S$ on the invariant energy $s' = w\alpha$
at large $s'$: $T_S$ decreases with $s'$ faster than $s'^{-h}$.

\section{Reducing the basic factorization to $k_T$ -factorization}

In order to bring the convolutions in Eqs.~(\ref{convs},\ref{convns})
to the form of the $k_T$ factorization, the integration over
$\alpha$ should be performed. Obviously, this
integration
cannot be performed in the straightforward way
without dealing with the
perturbative parts of the convolutions
because they depend on
$k^2$ and $k^2$ depends on $\alpha$ through Eq.~(\ref{sud}).
The only way to bring
these convolutions
to the form corresponding to the
$k_T$ -factorization
is to impose the following restriction
on the longitudinal component $w\alpha\beta$ of
$k^2$ compared to its transverse component:

\begin{equation}\label{kt}
w \alpha\beta \ll k_{\perp}^2~.
\end{equation}

Eq.~(\ref{kt}) allows us to disentangle the $\alpha$ and $\beta$ -dependence in
Eqs.~(\ref{convs},\ref{convns},\ref{convnsf},\ref{convsf}). Under
this condition, the upper (perturbative) blob in
Fig.~\ref{infraredfig1} approximately depends on $\beta$ and
$k_{\perp}$ whereas the lower blob approximately depends on
$\alpha$ and $k_{\perp}$. It allows us to perform the integration
over $\alpha$ in
Eqs.~(\ref{convs},\ref{convns},\ref{convnsf},\ref{convsf})
without involving the
perturbative components, which makes possible to bring
these equations
to the form corresponding to the $k_T$ -factorization.
To our knowledge, Eq.~(\ref{kt}) is not written
explicitly in the literature. However, we would like to stress that it
has commonly been used, though in an inexplicit way and in
other terms\footnote{For example, in the book \cite{dkt} the Sudakov variables $\alpha$ and $k^2_{\perp}$
are replaced by $m^2 = (p-k)^2$ and $k^2$ respectively.},  in order to obtain integral
representations of the DIS structure functions, including
DGLAP.
At $x \sim 1$, the
strong inequality in  Eq.~(\ref{kt})
should be replaced by the weaker condition $w \alpha\beta \lesssim k_{\perp}^2$.
Eq.~(\ref{kt}) perfectly agrees with  the well-known fact (but does not follow from it) that
the virtualities of the ladder partons are transverse.
Applying Eq.~(\ref{kt}) to $A_{NS}$ allows us to neglect
the
$\alpha$ -dependence in the infrared-sensitive terms in
$M_{NS}$ and to keep it in
 $T_{NS}$ only. Performing the integration over $\alpha$ in
 Eq.~(\ref{convns}), we arrive at

\begin{equation}\label{convpsi}
A_{NS} (q,p) = \int_{0}^{w} \frac{d k^2_{\perp}}{k^2_{\perp}} \int
\frac{d \beta}{\beta}
M_{NS} \left(\ln (w\beta/k^2_{\perp}), \ln (Q^2/k^2_{\perp})\right)
\Psi_{NS} (w\beta, k^2_{\perp}) ,
\end{equation}
with

\begin{equation}\label{psi}
\Psi_{NS} \left( w\beta, k_{\perp}^2\right) = k^2_{\perp} \int_{- \infty}^{\infty} d \alpha
\frac{B T_{NS} \left(w \alpha, w \alpha\beta + k_{\perp}^2 \right)}
{(w \alpha \beta + k_{\perp}^2)^2} \approx \int_{k^2_{\perp}/w}^{k^2_{\perp}/w \beta} d \alpha
T_{NS} (w\alpha, k^2_{\perp}).
\end{equation}

At $x \sim 1$, the upper limit $w$ of the $k_{\perp}^2$ -integration in Eq.~(\ref{convpsi})
can be replaced by
$Q^2$ as it takes place in DGLAP.

Taking the imaginary part converts Eq.~(\ref{convpsi}) into the following
approximate expression for
the non-singlet structure functions $f_{NS}$:
\begin{eqnarray}\label{convns1}
f_{NS}(x, Q^2) =  \int_0^{w}
\frac{d k_{\perp}^2}{k_{\perp}^2} \int_{\beta_0}^1 \frac{d \beta}{\beta}
 f^{(pert)}_{NS}\left( w\beta, Q^2, k_{\perp}^2 \right)
\Im \Psi_{NS} \left( w \beta , k_{\perp}^2\right)~,
\end{eqnarray}
with
\begin{equation}\label{betazero}
\beta_0 = x + k^2_{\perp}/w \approx \max [x, k^2_{\perp}/w]
\end{equation}
 and
\begin{equation}\label{phi}
\Im \Psi_{NS} \left( w \beta, k_{\perp}^2, m \right) \approx
\int_{k_{\perp}^2/w}^{k_{\perp}^2/w \beta} d \alpha
 \Im T_{NS} \left(w \alpha,  k_{\perp}^2, m \right) \equiv
 \Phi_{NS} (w \beta, k_{\perp}^2)
\end{equation}
where we have introduced the non-singlet unintegrated parton distributions
 $\Phi_{NS} (\beta,k_{\perp}^2 )$.  They correspond to the initial
parton densities in the case of the $k_{\perp}$ -factorization.
Now we can represent the non-singlet
structure functions in terms of convolutions of the perturbative (partonic)
and non-perturbative parts:

\begin{eqnarray}\label{convmu}
f_{NS}(x, Q^2) =  \int_0^{w}
 \frac{d k_{\perp}^2}{ k_{\perp}^2} \int_{\beta_0}^1 \frac{d \beta}{\beta}
 f^{(pert)}_{NS}\left( w\beta, Q^2, k_{\perp}^2 \right)
\Phi_{NS} \left( w \beta , k_{\perp}^2\right)~.
\end{eqnarray}

The structure of Eqs.~(\ref{convpsi}, \ref{convmu}) is quite similar to the
one in Ref.~\cite{catani} and corresponds to the $k_T$ -factorization.
Introducing $x_0 = k_{\perp}^2/w$ and a mass scale $m$ (for example, it can be
the hadron target mass)
allows us to write Eq.~(\ref{convmu})
in the following symmetrical form:
\begin{eqnarray}\label{convf}
f_{NS}(x, Q^2) =
 \int_0^{w}
 \frac{d k_{\perp}^2}{k_{\perp}^2} \int_{\beta_0}^1 \frac{d \beta}{\beta}
 f^{(pert)}_{NS}\left( x/\beta, Q^2/k_{\perp}^2 \right)
\Phi_{NS} \left( x_0/\beta, k_{\perp}^2/ m^2\right)~.
\end{eqnarray}

The attractive feature of Eq.~(\ref{convf}) is that it exhibits the remarkable symmetry between
the arguments
of $f^{(part)}_{NS}$ and $\Phi_{NS}$: $x \leftrightarrow x_0$,
$Q^2 \leftrightarrow k_{\perp}^2$ and $k_{\perp}^2 \leftrightarrow m^2$. Nevertheless,
let us remind that though $x$ and $Q^2$ are conventionally regarded as independent variables, strictly
speaking they are not independent. Indeed, they can be regarded as independent
only in the case when $Q^2$ is kept fixed while $w$ is scanned. In the opposite
case when $w$ is fixed and $Q^2$ is varied, the variables $Q^2$ and $x$ are not independent at all. So,
the parametrization of the structure functions in terms of the really independent variables
$w$ and $Q^2$ would be preferable at least for the theoretical analysis.

Similarly to the non-singlet case,  the singlet amplitude $A_S$ can be written as follows:
\begin{equation}\label{asinglet}
A_{S} (q,p) = \int_{0}^{w} \frac{d k^2_{\perp}}{k^2_{\perp}} \int
\frac{d \beta}{\beta} \left(\frac{w\beta}{k^2_{\perp}}\right)
M_{S} \left(\ln (w\beta/k^2_{\perp}), \ln (Q^2/k^2_{\perp})\right)
\Psi_{S} (w\beta, k^2_{\perp}) ,
\end{equation}
with
\begin{equation}\label{psisinglet}
\Psi_{S} \left( w\beta, k_{\perp}^2\right) = (k^2_{\perp})^2 \int_{- \infty}^{\infty} d \alpha
\frac{B T_{S} \left(w \alpha, w \alpha\beta + k_{\perp}^2 \right)}
{(w \alpha \beta + k_{\perp}^2)^3} \approx
\int_{k^2_{\perp}/w}^{k^2_{\perp}/w \beta} d \alpha
T_{S} (w\alpha, k^2_{\perp}).
\end{equation}

The unintegrated parton distribution $\Phi (w\beta, k^2_{\perp})$ can be defined similarly
to the non-singlet case:
 \begin{equation}\label{phis}
\Im \Psi_S \left( w \beta, k_{\perp}^2, m \right) \approx
\int_{k_{\perp}^2/w}^{k_{\perp}^2/w \beta} d \alpha
 \Im T_S \left(w \alpha,  k_{\perp}^2, m \right) \equiv
\Phi_S (w \beta, k_{\perp}^2) .
\end{equation}
Therefore, the singlet structure function $f_S$ (i.e. the DIS structure function $F_1^S$) is
\begin{eqnarray}\label{convmusinglet}
f_S(x, Q^2) =  \int_0^{w}
 \frac{d k_{\perp}^2}{ k_{\perp}^2} \int_{\beta_0}^1 \frac{d \beta}{\beta}
f^{(pert)}_S\left( w\beta, Q^2, k_{\perp}^2 \right)
\Phi_{S} \left( w \beta , k_{\perp}^2\right)~.
\end{eqnarray}

\section{Infrared behavior of parton distributions in the $k_T$-factorization}

The generalized parton distributions $T_{S,NS}$ (in the expressions the Compton amplitudes) and $\Phi_{S,NS}$
(in the expressions the structure functions) include unperturbative
contributions. When the $k_T$ -factorization is used,
there is no problem about UV divergences but the IR divergences at
small $k_{\perp}$ must be regulated. Below we study the impact of
the integrability conditions on the properties of the parton distributions
in the framework of the $k_T$ -factorization.


Let us first consider the $k_{\perp}$ behavior of the non-singlets.
The explicit expressions
for $M_{NS}$ can be
different, depending on the approach chosen to calculate it but
their IR-divergent contributions
in
$M_{NS}$
are always logarithmic:
\begin{equation}\label{apertir}
M_{NS} \sim \ln^n (w \beta/k_{\perp}^2),~~\ln^l (Q^2/k_{\perp}^2),
\end{equation}
 with $n,l = 1,2,..$.
 So, $k^2_{\perp} \ll w\beta,~Q^2$
in any perturbative approach. It means that the $k_{\perp}$ -integration
is free of ultraviolet divergences. However,
the infrared divergences can appear in Eqs.~(\ref{convpsi},\ref{convmu}). In order to avoid them,
we have to put a restriction on $T_{NS}$ at small $k_{\perp}$:
\begin{equation}\label{tnsk}
T_{NS} \sim (k_{\perp}^2)^{\eta}~
\end{equation}
where $\eta > 0$.
The main difference between  Eqs.~(\ref{asinglet},\ref{convmusinglet}) and the non-singlets
(\ref{convpsi},\ref{convmu})
is the factor $w\beta/k^2$ in Eqs.~(\ref{asinglet},\ref{convmusinglet}).
It changes the non-singlet restriction (\ref{tnsk}) for

\begin{equation}\label{tsk}
T_S \sim (k_{\perp}^2)^{1 + \eta}~
 \end{equation}
at small $k_{\perp}$.
Besides regulating the IR-divergences, Eqs.~(\ref{tnsk},\ref{tsk}) have
another important meaning for the $k_T$ factorization:
they allow to neglect the region of small $k_{\perp}$ in the  convolutions,
solving therefore the problem of the mass singularities\cite{masssing}.

The initial parton densities $\Phi$ contribute to the DIS structure functions
when the $k_{\perp}$ -factorization is used. In this case, $k_{\perp}^2/\beta$
is their invariant energy, so Eqs.~(\ref{tnsalpha}, \ref{tsalpha}),
with $\alpha$ replaced by $k^2_{\perp}/w \beta$, predict
the following dependence of the initial parton distributions on $\beta$:
\begin{equation}\label{phiktbetans}
\Phi_{NS} \sim \beta^h,
\end{equation}
\begin{equation}\label{phiktbetas}
\Phi_S \sim \beta^{-1 + h} .
\end{equation}

They mean that $\Phi_{NS}$ should be regular in
$\beta$ when $\beta \to 0$ and $\Phi_S$ can be singular but with $h < 1$. These
restrictions and the ones in Eqs.~(\ref{tnsk}, \ref{tsk})
must be respected when fits for the initial parton
distributions $\Phi_S, \Phi_{NS}$ are composed in the framework of the $k_T$ -factorization.

\section{Collinear factorization for the structure functions}

In this Sect. we consider the infrared properties of the DIS structure functions when
the collinear factorization is introduced. Historically, the collinear factorization\cite{fact}
was the first one considered in the literature.
In general, the basic assumption of the collinear factorization is to introduce the scale
$\mu$ for the parton virtualities $k^2_i$, with $k^2_i > \mu^2 > \Lambda^2_{QCD}$
as an explicit border between the perturbative (hard) and non-perturbative
(soft) domains of QCD.
For describing the DIS structure functions, this factorization first was used
to construct
the DGLAP evolution equations where
the ordering (\ref{dglapord}) was suggested to evolve the structure functions from
$\mu^2$ to $Q^2$. Later, the collinear factorization
combined with the ordering (\ref{dlord})
was used to describe some of the structure functions in the small-$x$
region where the total resummation of $\ln^k(1/x)$ becomes essential
(see Ref.~\cite{g1sum} and Refs therein for detail).

\subsection{Transition from the $k_T$ -factorization to the collinear factorization}

Eqs.~(\ref{convpsi},\ref{asinglet}) for the forward Compton amplitudes
as well as Eqs.~(\ref{convmu},\ref{convmusinglet}) for the structure functions
involve integrations over $\beta$ and $k_{\perp}$, whereas similar expressions in the
collinear factorization operate with integrations over $\beta$ only. So,
our aim now is to integrate over $k_{\perp}$ in
Eqs.~(\ref{convpsi},\ref{asinglet},\ref{convmu},\ref{convmusinglet}) without
dealing with the perturbative contributions. However,
both the perturbative and unperturbative factors in
Eqs.~(\ref{convpsi},\ref{asinglet},\ref{convmu},\ref{convmusinglet}) depend
on $k_{\perp}$. It makes impossible any straightforward $k_{\perp}$ -integration of
the unperturbative terms. Therefore, a transition from the $k_{\perp}$ -factorization
to the collinear one can be done only approximately. In a sense, it
is similar to the transition form the basic form of the factorization to the
$k_T$ -factorization we have considered in Sect.~VI.

\begin{figure}[h]
\begin{center}
\begin{picture}(240,180)
\put(0,0){\epsfxsize=50mm \epsfbox{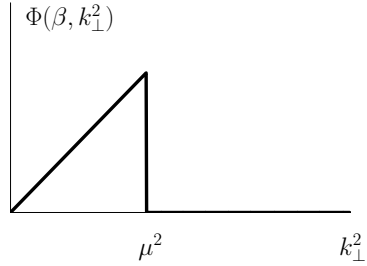} }
\end{picture}
\end{center}
\caption{\label {infraredfig5} The plot for
$\Phi(\beta,k_{\perp}^2)$ defined according to
Eq.~(\ref{phicolteta})}
\end{figure}

In order to do it, let us first  suppose that $\Phi_{S,NS} (\equiv \Phi)$
in Eqs.~(\ref{convmu},\ref{convmusinglet})
can be represented as
\begin{equation}\label{phicolteta}
\Phi = \varphi(\beta, k_{\perp}^2)~ \Theta (\mu^2 - k_{\perp}^2 ) ,
\end{equation}
with $\varphi(\beta, k_{\perp}^2)$ decreasing rapidly at large
$k_{\perp}$ and respecting the restrictions in
Eqs.~(\ref{tnsk},\ref{tsk}). This choice of $\Phi$ is depicted in
Fig.~\ref{infraredfig5}.

It means that $\mu$ is the strict border between the contributions
forming the upper and lower blobs in Fig.~\ref{infraredfig1}: the
perturbative corrections with $k_{\perp} > \mu$ are attributed to
the upper blob and participate in the orderings of
Eqs.~(\ref{dglapord}, \ref{dlord}) while the lower blob includes
the contributions with $k_{\perp} < \mu$.

\begin{figure}[h]
\begin{center}
\begin{picture}(240,100)
\put(0,0){\epsfxsize=50mm \epsfbox{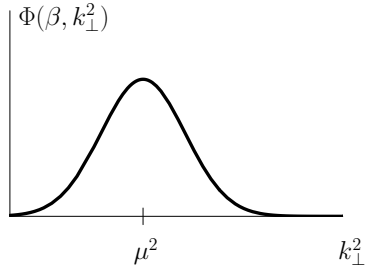} }
\end{picture}
\end{center}
\caption{\label {infraredfig6} The plot for
$\Phi(\beta,k_{\perp}^2)$ defined according to Eq.~(\ref{phicol})}
\end{figure}

Alternatively, the transition from the $k_T$- factorization to the
collinear factorization
can be done without using the $\Theta$-function. Let us assume
(see Fig.~\ref{infraredfig6}) that  $\Phi$ has a sharp maximum at
$k^2_{\perp} = \mu^2 \ll Q^2$ saturating the integration over
$k_{\perp}$ in Eqs.~(\ref{convmu},\ref{convf}).

For example, it takes place when the $k_{\perp}^2$- dependence of
$\Phi$ includes the Gaussian exponential (or a similar sharply
peaked factor). $\Phi (\beta, k_{\perp}^2)$ can also include a
flatter function of $k_{\perp}^2$:
\begin{equation}\label{phicol}
\Phi = \varphi(\beta, k_{\perp}^2) \exp \left[-\sigma \left(k_{\perp}^2 - \mu^2
\right)^2\right]
\end{equation}
where the function $\varphi (\beta, k_{\perp}^2)$ in
Eq.~(\ref{phicol}) respects Eqs.~(\ref{tnsk},\ref{tsk}). In this
case $\mu$ cannot be regarded as the border between the
contributions to the upper and lower blobs in
Fig.~\ref{infraredfig1}: $\Phi$ in Eq.~(\ref{phicol}) comprises
the contributions with arbitrary $k_{\perp}$.

Substituting Eq.~(\ref{phicolteta}) or (\ref{phicol}) in
Eqs.~(\ref{convpsi},\ref{asinglet},\ref{convmu},\ref{convmusinglet})
allows us to integrate $\Phi$ over $k_{\perp}^2$ independently of $f^{(pert)}$,
arriving at the standard
expressions with the collinear factorization.
In particular, combining (\ref{phicolteta}) or (\ref{phicol}) with  Eqs.~(\ref{convf}, \ref{convmusinglet})
we arise at the following generic expression for the structure functions:

\begin{equation}\label{convcol}
f(x, Q^2) = \sum_{r = q,g}
 \int_{\beta_0 (\mu)}^1 \frac{d \beta}{\beta}  ~f^{(pert)}_r \left( x/\beta, Q^2/\mu^2 \right)
\varphi^{(r)}(\beta, \mu^2)~
\end{equation}
where $\beta_0 (\mu) = x+\mu^2/w \approx x$. We have used in
Eq.~(\ref{convcol}) the generic notations $f,~f^{(pert)}_r$ both for the singlets and
non-singlets;  the subscripts $q,g$ in Eq.~(\ref{convcol}) refer to
the quark and gluon respectively.
The perturbative contributions $f^{(pert)}_r$ in
Eq.~(\ref{convcol}) are usually calculated either
in the framework of DGLAP or with
total resummation of the leading (and sub-leading) logarithms. The
transition from the $k_T$-factorization to the collinear factorization
does not affect the $\beta$-dependence in
Eqs.~(\ref{phiktbetans},\ref{phiktbetas}), so the singlet
$\varphi_S(\beta, \mu^2)$ and non-singlet $\varphi_{NS}(\beta, \mu^2)$
parton distributions used
in the framework of the collinear factorization should obey
the following relations respectively:
\begin{equation}\label{phicolns}
\varphi_{NS}(\beta, \mu^2) \sim \beta^h,
\end{equation}

\begin{equation}\label{phicols}
\varphi_S(\beta, \mu^2) \sim \beta^{-1 + h},
\end{equation}

with $h > 0$. We would like to stress here that the restrictions (\ref{phicolns}, \ref{phicols})
are not related to the integrability of the collinear factorization
convolution in Eq.~(\ref{convcol}) in itself. Indeed, the integral in Eq.~(\ref{convcol})
is convergent at any $\varphi$ because the integration over $\beta$ here
 runs over the compact region $[\beta_0 (\mu); 1]$. However,
 violations of (\ref{phicolns}, \ref{phicols}) would destroy
 the integrability of the more general factorization convolutions  in
Eqs.~(\ref{convs},\ref{convns}). We remind that in the present paper we have derived
(\ref{phicolns}, \ref{phicols})
from Eqs.~(\ref{convs},\ref{convns}).

Let us notice that $\mu^2$ in Eq.~(\ref{convcol}) stands both for the factorization scale and for the
starting point of the perturbative $Q^2$-evolution. On the other hand, this evolution
can start at an arbitrary point. When it starts at
$Q^2 = \mu^2_0$ (with $\mu^2_0 > \mu^2$),   Eq.~(\ref{convcol}) can be written as

\begin{equation}\label{convcolq}
f(x, Q^2) = \sum_{r = q,g}
 \int_{\beta_0 (\mu)}^1 \frac{d \beta}{\beta}  ~f^{(pert)}_r \left( x/\beta, Q^2/\mu^2_0 \right)
\widetilde{\varphi}^{(r)}(\beta, \mu^2_0)~,
\end{equation}
with $\widetilde{\varphi}^{(r)}(\beta, \mu^2_0)$ being new parton distributions. They are defined
at the new factorization scale $\mu^2_0$. The values of $\mu^2_0$ are located within the range of the Perturbative
QCD, so the parton distributions $\widetilde{\varphi}^{(r)}(\beta, \mu^2_0)$
include both non-perturbative and perturbative contributions.
 Although both $f^{(pert)}_r \left( x/\beta, Q^2/\mu^2_0 \right)$ and
$\widetilde{\varphi}^{(r)}(\beta, \mu^2_0)$
depend on $\mu^2_0$, the integral in Eq.~(\ref{convcolq}) is obviously
$\mu^2_0$-independent exactly as it is
independent of contributions of any intermediate integration point.
Indeed,
the inclusion of the perturbative contributions into
$\widetilde{\varphi}(x, \mu^2_0)$ implies that there is another scale
$\mu$ (with $\mu < \mu_0$) in $\widetilde{\varphi}(x, \mu^2_0)$,
so the perturbative methods
can be used for the evolution from $\mu^2$ to $\mu^2_0$. Obviously, the transition from
$\varphi_{S,NS}(x, \mu^2)$ to $\widetilde{\varphi}_{S,NS}(x, \mu^2_0)$ can be
done with the same technique as the evolution from $\mu^2_0$ to $Q^2$ in
$f^{(pert)}$:

\begin{equation}\label{evol}
f (Q^2) = U(Q^2, \mu^2_0) \widetilde{\varphi} \otimes (\mu^2_0), ~~
\widetilde{\varphi} (\mu^2_0) = U (\mu^2_0, \mu^2) \otimes \varphi (\mu^2).
\end{equation}
where $U (b,a)$ is a generic notation for the evolution operator, with
$a (b)$ standing for the starting (final) point of the evolution. We have
skipped dependence on all variables in Eq.~(\ref{evol}) inessential at the moment.
 For instance, in the case of the LO DGLAP $\widetilde{\varphi} (\mu^2_0)$
 and $\varphi (\mu^2)$ are related in the momentum space as follows:

\begin{equation}\label{phimuq}
\widetilde{\varphi}(\omega, \mu^2_0) = \left[\frac{\ln \left(\mu^2_0/\Lambda^2\right)}
{\ln \left(\mu^2/\Lambda^2\right)}\right]^{\gamma^{(1)}(\omega)/b} \varphi (\omega, \mu^2),
\end{equation}
where $\gamma^{(1)} (\omega)$ is the LO anomalous dimension and
$b= [33 - 2 n_f]/(12 \pi)$ is the first coefficient of the $\beta$ -function.
In this case the perturbative structure functions $f^{(pert)} (\omega, Q^2)$
and $f^{(pert)} (\omega, \mu^2_0)$ are related similarly:

\begin{equation}\label{fpertq}
f^{(pert)} (\omega, Q^2) =
\left[\frac{\ln \left(Q^2/\Lambda^2\right)}
{\ln \left(\mu^2_0/\Lambda^2\right)}\right]^{\gamma^{(1)}(\omega)/b}
f^{(pert)} (\omega, \mu^2_0)
\end{equation}

Substituting Eqs.~(\ref{phimuq},\ref{fpertq}) into (\ref{convcolq})
kills the dependence of $f(x,Q^2)$ on the arbitrary scale $\mu^2_0$ but
does not affect its dependence on the scale $\mu$. We remind we
have defined the scale
$\mu$ in Eqs.~(\ref{phicolteta}, \ref{phicol}) as the scale separating
the perturbative and non-perturbative contributions.


Our restrictions (\ref{phicolns},\ref{phicols}) look invalid in the
framework of the approach suggested in Ref.~\cite{efr}. We will address it
as the EGR-approach. In this approach the factorization scale
$\mu_0$ can also be chosen arbitrary and the QCD coupling is kept fixed.
The EGR-approach is based on the  observation that
the lader Feynman graphs contributing to the Compton amplitude yield the
sub-leading IR-stable logarithmic contributions in the
kinematic region where the virtualities $k^2_r$ ($r=1,2,..$) of the vertical partons
are very small:
\begin{equation}\label{efrreg}
0 < |k^2_r| < \mu^2_0 ,
\end{equation}
while virtualities of the horizontal partons are greater than $\mu^2_0$.
Like in Eqs.~(\ref{dglapord},\ref{dlord}),
we have numerated  the ladder momenta $k_r$ in (\ref{efrreg}) from the bottom
to the top of the ladders graphs.
Integrations of the ladder graphs over $k_r$ in the region
 (\ref{efrreg}) yield logarithmic
contributions $\sim \ln^n x$ for the non-singlets and
$\sim (1/x) \ln^n x$ for the singlet. Resummation of them leads to the Regge factors $\sim x^{-a}$
for the non-singlets and $\sim x^{-1-a}$ for the singlet (see Refs.~\cite{efr,wu} for
detail).
As $k^2_r$ in the region (\ref{efrreg}) are small, these Regge factors,
according to  Ref.~\cite{efr} should
be  included
into $\widetilde{\varphi}(x, \mu^2_0)$
.
The sub-leading logarithmic contributions $\sim \ln^n x$ are IR-stable, so
the EGR-approach
operates with the scale $\mu_0$ only and does not need any
additional scale $\mu$.
This approach obviously contradicts our
approach and makes invalid our restrictions (\ref{phicolns},\ref{phicols}).
We plan to compare our and EGR approaches in detail in our forthcoming paper.
Here we just mention that the EGR- approach leads to a 
 much more complicated factorization construction 
than the one depicted in Fig.~1: 
instead of 
two blobs 
it involves a pile (etagere) of perturbative and non-perturbative
blobs.

\subsection{Restrictions on the singular factors in the DGLAP fits}

Now let us focus on the convolution in Eq.~(\ref{convcol}), with the DGLAP expression
used for the perturbative
structure functions $f^{(pert)}_r$. In the standard notations, Eqs.~(\ref{convcol})
takes the following form:

\begin{equation}\label{convdglap}
f(x, Q^2) = \sum_{r = q,g}
 \int_{x}^1 \frac{d y}{y}  ~f^{(r)}_{DGLAP}\left( x/y, Q^2/\mu^2 \right)~
\delta r (y, \mu^2)~.
\end{equation}
where
we have dropped the term $\mu^2/w$ in the lowest integration limit.
The initial parton densities $\delta q (x), \delta g(x)$ in Eq.~(\ref{convdglap}) are
obtained by fitting the experimental data at $\mu \sim 1$ GeV. They include
both singular factors $x^{-a}$ (with $a > 0$) and  regular factors $\sim (1-x)^b,
(1 + cx^d)$, with $b,c,d > 0$.

Comparing Eqs.~(\ref{convcol}) and (\ref{phicolns},\ref{phicols}) with (\ref{convdglap})
drives us to conclude that the standard DGLAP fits with
the singular factors $x^{-a}$ for the non-singlet structure functions are excluded by
Eq.~(\ref{phicolns}). On the other hand, Eq.~(\ref{phicols}) admits the use of the
singular factors for the singlet $F_1$, however with the exponent $a <1$.
We would like to stress here that the use of the singular factors cannot be
excluded by requirements on the integrability of
the structure functions (where all integration limits are finite).
However, the use of such factors contradicts to
the integrability the expressions for the forward Compton amplitudes where the integrations
run over the whole phase space.
Now let us comment on these
results in more detail and consider below both the $k_T$- and collinear
factorizations.

First of all,
we infer that the fits for the polarized quark and gluon  distributions
should satisfy the restriction (\ref{phiktbetans}) or (\ref{phicolns}), depending on
whether the $k_T$- or collinear factorization is used.

Then,
the spin-independent non-singlet structure functions involve
the unpolarized quark distributions only, so each of the quark distributions should
again satisfy
Eqs.~(\ref{phiktbetans},\ref{phicolns}). It is true also for the singlet $F_2$.

The difference between the restrictions (\ref{phiktbetans},\ref{phicolns}) for the non-singlets
and the ones (\ref{phiktbetas},\ref{phicols}) for the singlets
arises entirely
from the perturbative components
of the singlet $F_1$ $(\equiv F_1^S)$ and
the other structure functions.
$F_1^S$  involves
the unpolarized quark and gluon distributions. The quark distributions in $F_1^S$
are the same as the ones to the non-singlets $F_1$, so they
satisfy Eqs.~(\ref{phiktbetans},\ref{phicolns}) but the
restriction on the gluon
distribution to $F_1^S$ is softer because it is given by Eqs.~(\ref{phicolns},\ref{phicols}).
On the other hand, the same gluon distribution is supposed to contribute to the
singlet $F_2$ and therefore it should also satisfy
Eqs.~(\ref{phiktbetans}, \ref{phicolns}).
Eventually, we conclude that the singular factors $x^{-a}$ should be absent
in all parton distributions.

In Refs.~\cite{egtinp, g1sum} we showed
that the singular factors in the DGLAP -fits were introduced to match experimental data at small $x$.
 The reason for introducing them is that they cause a steep rise of the structure functions at small $x$, thereby
making possible the use of DGLAP in the small-$x$ region.
As a matter of fact, they mimic the total resummation of the leading logarithms of $x$,
 which is beyond the reach of DGLAP.
With the resummation taken into account, such a rise of the structure functions at small $x$ is
achieved automatically,
so the singular factors become unnecessary. In addition, we have shown
in the present paper that
there are theoretical restrictions on the singular factors
following from the requirement of integrability of
the factorization convolutions for the forward Compton amplitudes.

\section{Summary}

In the present paper we have derived the $k_T$- and collinear factorizations for the
Compton amplitude and DIS structure functions and obtained the restrictions on the
parton distributions to the DIS structure functions. The
parton distributions available in the
literature are fixed
phenomenologically. In contrast,  we have
obtained the theoretical restrictions on them,  exploiting the obvious mathematical requirement of
integrability of the factorization convolutions.

We began by considering  the convolutions (\ref{convaqg}) depicted
in Fig~1 where the amplitude of the forward Compton scattering off
a hadron target is represented through two blobs connected by the
two-parton intermediate states involving either quarks or gluons.
Strictly speaking, the convolutions in Eq.~(\ref{convaqg}) do not
correspond to the conventional concept of the QCD factorization
because each of the amplitudes there (each blob in
Fig.~\ref{infraredfig1}) can contain the both non-perturbative and
perturbative contributions, so we called them the primary
convolutions and considered their integrability in the Born
approximation. Accounting for the radiative corrections converted
the Born amplitudes into Eqs.~(\ref{convs},\ref{convns}). These
convolutions complemented by the restrictions in
Eqs.~(\ref{tnsk2}, \ref{tsk2}) rigorously correspond to the QCD
factorization concept: the upper blob in Fig.~\ref{infraredfig1}
becomes free of non-perturbative contributions while the lower
blob is non-perturbative. However,
the factorization convolutions in Eqs.~(\ref{convs},\ref{convns})
differ from the convolutions in the $k_T$- and collinear
factorization and involve
the totally
unintegrated parton distributions, so we call
such a factorization the basic
factorization.  The convolutions in
Eqs.~(\ref{convs},\ref{convns}) are UV-stable when the
restrictions (\ref{tnsalpha},\ref{tsalpha}) are fulfilled.
Applying the Optical theorem to Eqs.~(\ref{convns},\ref{convs}),
we arrived at the expressions (\ref{convnsf},\ref{convsf}) for the
singlet and non-singlet structure functions in the basic
(totally
unintegrated) form.

We showed that reducing the basic factorization to the $k_T$- factorization
can be done only when the approximation in Eq.~(\ref{kt}) was accepted.
Using this result, we obtained
Eqs.~(\ref{convmu},\ref{convmusinglet}) for the structure
functions in the framework of the $k_T$ -factorization.
Investigating the impact of the integrability restrictions
(\ref{tnsalpha},\ref{tsalpha}) on (\ref{convmu},\ref{convmusinglet}), we obtained
the theoretical
restrictions on the parton distributions at the
the $k_T$ -factorization: Eqs.~(\ref{phiktbetans},\ref{phiktbetas})
show that the non-singlet
parton distributions should be regular in the longitudinal momentum
$\beta$ while the distributions for the singlet $F_1$ can include
the singular factors $\beta^{-h}$,
but with $h < 1$. Besides, the parton distributions should be regular in the transverse
momenta. These restrictions should be
respected in fits for the parton distributions
composed in the framework of
the $k_T$ -factorization.
When the $k_{\perp}$-dependence of those parton distributions exhibits  a sharp maximum
at $k_{\perp}^2 = \mu^2$, as shown in Figs.~5,6,
the $k_T$ -factorization can be reduced to the collinear
factorization, with $\mu$ playing the role of the factorization scale. This
our treatment of $\mu$ can be checked by fitting experimental data in
the framework of the $k_T$- factorization.

The integrability of
the forward Compton scattering amplitudes at the basic factorization
leads to the
restrictions (\ref{phicolns},\ref{phicols}) on the singular factors $x^{-a}$
 in the fits for the initial parton distributions:
$a < 1$ in the fits for the singlet $F_1$
and $a \leqslant 0$ in the fits for other structure functions providing the perturbative
corrections are attributed to the upper blob.

There is no essential difference between the restrictions
in Eqs.~(\ref{phicolns},\ref{phicols}) and (\ref{phiktbetans},\ref{phiktbetas})
obtained for the collinear and $k_T$- factorizations respectively.
On the other hand, the restrictions
(\ref{phiktbetas},\ref{phicols}) on
the unpolarized singlet differ from the restrictions (\ref{phiktbetans},\ref{phicolns})
for all other parton distributions. However, this difference
results altogether from the peculiar ($\sim 1/x$) perturbative $x$-evolution
 of $F_1$ singlet
compared to the other structure functions,
i.e. from the difference between singlet and non-singlet coefficient functions. So, as
the same parton distributions contribute to the singlet $F_1$ and the other
spin-independent structure functions,
all the parton distributions should satisfy Eqs.~(\ref{phiktbetans},\ref{phicolns}).
The same is true for the polarized parton distributions.
The necessity to keep the singular factors in the fits should be regarded as a clear
indication that the perturbative components of the structure functions used in the
analysis lack the resummation of the contributions $\sim \ln^k (1/x)$.

To conclude, let us notice that the results for
the non-perturbative contributions to the Compton amplitude and DIS structure functions
obtained in the present paper
follow from general properties of the primary convolutions in Eq.~(\ref{convaqg})
and because of that they can easily be applied to other inclusive hadronic processes at high energies.

\section{Acknowledgements}

We are grateful to A.V.~Efremov, M.~Anselmino, L.~Magnea and I.F.~Ginzburg for useful discussions. The work is partly supported by Grant RAS 9C237,
Russian State Grant for Scientific School
RSGSS-65751.2010.2 and EU Marie-Curie Research Training
Network under contract MRTN-CT-2006-035505 (HEPTOOLS).


\appendix
\section{Simplification of $B_{\mu\nu}$}

The tensor $\hat{k} A^{(q~B)}_{\mu\nu} \hat{k}$ in Eq.~(\ref{kaqbornk}) depends on
 the Lorentz
subscripts $\mu$ and $\nu$ through the term
\begin{equation}\label{bmunu}
B_{\mu\nu} \equiv \hat{k}\gamma_{\nu}(\hat{q}+ \hat{k})\gamma_{\mu}\hat{k} .
\end{equation}
$B_{\mu\nu}$ contains contributions symmetrical and antisymmetrical with respect  $\mu$ and $\nu$: $B_{\mu\nu} = B_{\mu\nu}^{(sym)} + B_{\mu\nu}^{(asym)}$ .
Let us consider first the antisymmetrical part, $B_{\mu\nu}^{(asym)}$. It contributes to the spin-dependent part
$\widetilde{A}^{(spin)}_{\mu\nu}$
of the amplitude $\widetilde{A}^{(pert)}_{\mu\nu}$.  $B_{\mu\nu}^{(asym)}$ can be simplified as follows:
\begin{equation}\label{basym}
B_{\mu\nu}^{(asym)} = -\imath (1 + \alpha) \epsilon_{\mu\nu\lambda\rho} q_{\lambda}
[\gamma_5 \hat{k} \gamma_{\rho} \hat{k}]
.
\end{equation}
The symmetrical part of $B_{\mu\nu}$ can be simplified similarly. For example, the contribution
in $B_{\mu\nu}^{(sym)}$ proportional to $g_{\mu\nu}$ is
\begin{equation}\label{bmunusym}
g_{\mu\nu} (1+ \alpha) q_{\rho}[\hat{k} \gamma_{\rho} \hat{k}] .
\end{equation}
Now let us simplify $\hat{k} \gamma_{\rho} \hat{k}$.

Using the Sudakov parametrization $k = \alpha q + \beta p + k_{\perp}$ in the c.m.f. where
 $p = \sqrt{w/2}(1,0,0,1)$ and $q = \sqrt{w/2}(1,0,0,-1)$,
we rewrite $\hat{k} \gamma_{\rho} \hat{k}$ as follows:
\begin{eqnarray}\label{kklt}
\hat{k}_{\perp} \gamma_{\rho} \hat{k}_{\perp} &=&
(\hat{k}_{\perp} \gamma_{\rho} \hat{k}_{\perp})^L +
(\hat{k}_{\perp} \gamma_{\rho} \hat{k}_{\perp})^T, \\ \nonumber
(\hat{k}_{\perp} \gamma_{\rho} \hat{k}_{\perp})^L &=&
(\alpha \hat{q} + \beta \hat{p}) \gamma_{\rho}
(\alpha \hat{q} + \beta \hat{p}),  \\ \nonumber
(\hat{k}_{\perp} \gamma_{\rho} \hat{k}_{\perp})^T
 &=& \hat{k}_{\perp} \gamma_{\rho} \hat{k}_{\perp}
\end{eqnarray}

As the subscript $\rho$ runs in the longitudinal space $(\rho = 0,3)$,
\begin{equation}\label{kkt}
(\hat{k} \gamma_{\rho} \hat{k})^T =  k^2_{\perp} \gamma_{\rho} .
\end{equation}
This is the only important term in LLA. When we go beyond the LLA, the longitudinal
term cam also be essential. In order to  deal with it, we represent $\gamma_{\rho}$
in terms of $\hat{q}, \hat{p}$:
\begin{equation}\label{gpq}
\gamma_0 = \frac{\hat{p} + \hat{q}}{\sqrt{w}},~~\gamma_3 = \frac{\hat{p} - \hat{q}}{\sqrt{w}}
\end{equation}
Substituting it into Eq.~(\ref{kklt}) and neglecting terms $ \sim q^2, p^2$, we obtain
\begin{eqnarray}\label{kkl}
(\hat{k}_{\perp} \gamma_{\rho} \hat{k}_{\perp})^L &=&
\frac{\alpha^2  \hat{q}\hat{p}\hat{q} +
\beta^2 \hat{p}\hat{q}\hat{p}}{\sqrt{w}} = w[(\alpha^2 + \beta^2) \gamma_0 +
(-\alpha^2 + \beta^2) \gamma_3] , \\ \nonumber
(\hat{k}_{\perp} \gamma_{\rho} \hat{k}_{\perp})^L &=&
\frac{\alpha^2  \hat{q}\hat{p}\hat{q} -
\beta^2 \hat{p}\hat{q}\hat{p}}{\sqrt{w}} = w[(\alpha^2 - \beta^2) \gamma_0 +
(\alpha^2 + \beta^2) \gamma_3] .
\end{eqnarray}

The contributions $\sim \beta^2$ in Eq.~(\ref{kkl}) can be dropped when Leading Logarithmic
Approximation is used to calculate $A^{(pert)}$. It follows form Eqs.~(\ref{kklt}, \ref{kkl})
that  $\hat{k} \gamma_{\rho} \hat{k} \sim \alpha^2$ when $\alpha \to \infty$. On the contrary,
when $\alpha$ is small, $\hat{k} \gamma_{\rho} \hat{k} = k^2_{\perp} \gamma_{\rho}$ .
It is convenient to use the interpolation expression for $\hat{k} \gamma_{\rho} \hat{k}$:
\begin{equation}\label{kkint}
\hat{k} \gamma_{\rho} \hat{k} \approx \gamma_{\rho} [w(\alpha^2 + \beta^2) + k^2_{\perp}]
\equiv \gamma_{\rho} B .
\end{equation}
It leads to the following approximative expressions for the symmetrical and antisymmetrical parts of
$B_{\mu\nu}$:
\begin{eqnarray}\label{bmunuint}
B_{\mu\nu}^{(sym)} &=& g_{\mu\nu} \hat{q} B , \\ \nonumber
B_{\mu\nu}^{(asym)} &=& \imath \epsilon_{\mu\nu\lambda\rho} q_{\lambda} \gamma_{\rho} B
\end{eqnarray}

\section{Projection operators for the Compton amplitude}

 The standard way of studying $A_{\mu\nu}$
is to introduce the projection operators and expanding $A_{\mu\nu}$ into a series of invariant amplitudes.
The spin-independent part $A^{(unpol)}_{\mu\nu}$ of $A_{\mu\nu}$ is parameterized as follows:
\begin{equation}
\label{aunpol}
A^{(unpol)}_{\mu\nu} = P^{(1)}_{\mu\nu} A_1 +
P^{(2)}_{\mu\nu} A_2
\end{equation}
where
\begin{equation}\label{punpol}
P^{(1)}_{\mu\nu}= \left( -g_{\mu\nu} + \frac{q_{\mu} q_{\nu}}{q^2} \right),~
P^{(2)}_{\mu\nu} = \left[P_{\mu} - q_{\mu} \frac{pq}{q^2}\right]\left[P_{\nu} - q_{\nu} \frac{pq}{q^2}\right]
\frac{1}{pq}
\end{equation}
while the expansion of the spin-dependent part $A^{(spin)}_{\mu\nu}$ is
\begin{equation}
\label{aspin} A^{(spin)}_{\mu\nu} = P^{(3)}_{\mu\nu} A_3 +
P^{(4)}_{\mu\nu} A_4,
\end{equation}
with
\begin{equation}\label{pspin}
P^{(3)}_{\mu\nu} = \imath \epsilon_{\mu\nu\lambda\rho} q_{\lambda} S_{\rho} \frac{m}{pq} ,~
P^{(4)}_{\mu\nu} = \imath \epsilon_{\mu\nu\lambda\rho} q_{\lambda} S_{\rho}
\frac{m}{pq}
\left[S_{\rho} - p_{\rho}\frac{qS}{pq}\right] ~.
\end{equation}
Any of the Compton
invariant amplitudes $A_{1,2,3,4}$ includes
both the flavor singlet and non-singlet  contributions. Their imaginary parts
correspond to the singlet and non-singlet components of the DIS structure functions.

\end{document}